\documentclass{ptephy_v1}
\begin{document}
\def\RR{{}^{\scriptscriptstyle(3)}\!R}
\def\v#1{\mbox{\boldmath{$#1$}}}
\def\aD{a_{\scriptscriptstyle D}}

\title{A possible solution to the Hubble constant discrepancy
  --- Cosmology where the local volume
  expansion is driven by the domain average density ---} 

\author{Masumi Kasai}
\affil{Graduate School of Science and Technology,
  Hirosaki University, Hirosaki, Aomori 036-8561,
  Japan\email{kasai@hirosaki-u.ac.jp}} 

\author[2]{Toshifumi Futamase}
\affil{Faculty of Science, Kyoto Sangyo University, Kyoto 603-8555,
  Japan\email{tof@cc.kyoto-su.ac.jp}}

\begin{abstract}
The Hubble constant problem is the discrepancy 
between different measurements of the Hubble constant in different scales.
We show that this problem can be resolved within the
general relativistic framework of the perturbation theory in
the inhomogeneous universe, with the help of spatial averaging
procedure over a finite local domain in the $t=\mbox{const.}$
hypersurface.
The idea presented in this paper is unique in the sense that it has all of the following properties. 
a) It is based on the general relativistic perturbation theory, with
ordinary dust matter only. No strange matter nor energy components are required. 
b) The employment of the spatially invariant averaging procedure on the finite domain  is essential. 
c) The key is the first-order effect of the inhomogeneities in the
linear perturbation theory. No non-linear effects are required.
\end{abstract}

\subjectindex{E60, E56}

\maketitle

\section{Introduction}
Recent high precision measurements of the Hubble constant $H_0$ show the
large discrepancy.  The Planck team's value for $H_0$ was $67.4\pm 0.5
\mbox{km/s/Mpc}$ \cite{1}, which was reported from the Planck satellite observing
the cosmic microwave background at very distant and large scale.
On the other hand, the Supernova $H_0$ for the Equation of State
(SH0ES) Collaboration reported a $H_0$ value $73.24\pm
1.74\mbox{km/s/Mpc}$ \cite{2}, which was based on measurements of the supernovae
in our cosmic neighborhood.
The result differs from Planck's by more than $3 \sigma$, a highly
statistically significant discrepancy which could not be easily
explained.

We show that the Hubble constant problem, the discrepancy between the
measurements of $H_0$ in different scales, can be resolved within the
general relativistic framework of the perturbation theory in the
inhomogeneous universe, with the help of the three-dimensional averaging
procedure over a finite domain in the $t=\mbox{const.}$
hypersurface.  

The idea presented in this paper is unique in the sense that it has
all of the following properties.
\begin{itemize}
\item[a)] It is based on the general relativistic perturbation theory, with
ordinary dust matter only. No strange matter nor energy components are
required. 
\item[b)] The employment of the spatially invariant averaging
  procedure on the finite domain  is essential. 
\item[c)] The key is the first-order effect of the inhomogeneities in the
linear perturbation theory. No non-linear effects are required.
\end{itemize}

\section{Basic equations}
In this section, we briefly summarize the basic equations \cite{3,4}.
We use the following convention:
Greek indices $\mu, \nu, \dots$ run from 0 to 3, Latin indices $i, j,
k, \dots$ run from 1 to 3, and the speed of light is unity, $c=1$.

We consider the model which contains irrotational dust with density
$\rho$ and four-velocity $u^{\mu}$.
In comoving synchronous gauge, which we adopt throughout the paper,
$u^{\mu} = (1,0,0,0)$ and the line
element can be written in the form
\begin{equation}
  ds^2 = -dt^2 + g_{ij} dx^i dx^j. 
\end{equation}
The Einstein equations read
\begin{equation}
  \label{eq:G00}
  \frac{1}{2}\left\{
    (K^i_{\ i})^2 - K^i_{\ j} K^j_{\ i} +
    \RR^i_{\ i} 
    \right\} = 8\pi G \rho, 
\end{equation}
\begin{equation}
  \label{eq:G0i}
  K^j_{\ i|j} - K^j_{\ j|i} = 0, 
\end{equation}
\begin{equation}
  \label{eq:Rij}
  \dot{K}^i_{\ j} + K^k_{\ k} K^i_{\ j} +
  \RR^i_{\ j} = 4\pi G \rho \delta^i_{\ j},
\end{equation}
where an overdot denotes $\partial/\partial t$, $|$ denotes the
three-dimensional covariant derivative with respect to $g_{ij}$, 
\begin{equation}
  K^i_{\ j} \equiv \frac{1}{2}g^{ik}\dot{g}_{kj}
\end{equation}
is the extrinsic curvature, and $\RR^i_{\ j}$ is the Ricci tensor of
the three-dimensional space with the spatial metric $g_{ij}$. 

The energy equation is

\begin{equation}
  \label{eq:ene}
  \dot{\rho} + K^{i}_{\ i} \,\rho = 0. 
\end{equation}

\section{The homogeneous and isotropic ``background'' }

The homogeneous and isotropic ``background'' is characterized by
the isotropic expansion:
\begin{equation}
  K^i_{\ j} = \frac{\dot{a}}{a} \delta^i_{\ j}, 
\end{equation}
where $a = a(t)$ is the scale factor.  Then, the Einstein equations
and the energy equation require that the three-dimensional space is of
constant curvature with the curvature constant $K$, i.e.,
\begin{equation}
  \RR^i_{\ j} = 2 \frac{K}{a^2} \delta^i_{\ j}, 
\end{equation}
and the density distribution is homogeneous, i.e., $\rho = \rho_b(t)$. 

Then, the Einstein equation Eq. (\ref{eq:G00}) and the energy
equation Eq. (\ref{eq:ene}) for the background 
are
\begin{equation}
  \left(\frac{\dot{a}}{a}\right)^2 + \frac{K}{a^2} = \frac{8\pi G}{3}
  \rho_b, 
\end{equation}
\begin{equation}
  \dot{\rho}_b + 3\frac{\dot{a}}{a} \rho_b = 0. 
\end{equation}
For the sake of simplicity, hereafter, we restrict ourselves to the
case of
$K=0$ background.
Generalizations to $K\neq 0$ background cases are straightforward.

\section{Weakly perturbed inhomogeneous universe}

The universe in reality is neither perfectly homogeneous nor
isotropic.  We assume that the inhomogeneities are small and briefly
summarize the results of linear perturbation theory, only considering
the scalar perturbations.  We can express the metric and the energy
density in the perturbed universe as follows:
\begin{equation}
  ds^2 = -dt^2 + a^2 \left(\delta_{ij} + 2 E_{,ij} + 2
    F\delta_{ij}\right) dx^i dx^j, 
\end{equation}
\begin{equation}
  \rho = \rho_b(1 + \delta). 
\end{equation}
From the linearized Einstein equations and the energy equation,
we obtain the second-order differentiation equation for the density
contrast $\delta$: 
\begin{equation}\label{eq:2delta}
  \ddot{\delta} + 2\frac{\dot{a}}{a} \dot{\delta} - 4 \pi G \rho_b
  \delta = 0.
\end{equation}
Under the normalization $a(t_0) =1$ at the present time $t_0$ and
neglecting the decaying mode solution,  the growing mode solution
for  $\delta$, which is proportional to 
$a(t)$ in the 
$K=0$ background, 
can be written as
\begin{equation}\label{eq:linisol0}
  \delta = \frac{2}{3H_0^2} a(t) \Delta \phi(\v{x}), 
\end{equation}
where $\Delta \equiv \delta^{ij} \partial_i \partial_j$ is the
Laplace operator.
The function $\phi(\v{x})$ does not depend on $t$ and can
be regarded as the Newtonian potential at the present time $t_0$
in the sense
\begin{equation}\label{eq:linisol1}
  \Delta \phi(\v{x}) =
  \frac{3 H_0^2}{2 a(t_0)} \delta(t_0,\v{x}) = 4\pi
    G\rho_b(t_0)\delta(t_0, \v{x}). 
  \end{equation}

Using $\phi(\v{x})$, the solutions for the metric linear perturbations can be
written as
\begin{equation}
  E=-\frac{2a(t)}{3H^2_0} \phi(\v{x}), \quad F = -\frac{5}{3}\phi(\v{x}),
\end{equation}
therefore, the line element is
\begin{equation}\label{eq:linisol2}
  ds^2 = -dt^2 + a^2\left( \delta_{ij}
    - \frac{4 a(t)}{3 H_0^2} \phi(\v{x})_{,ij}
    -\frac{10}{3} \phi(\v{x})\delta_{ij}\right) dx^i dx^j. 
\end{equation}

\section{The average density  and the volume
  expansion of a finite domain $D$}

In the previous section, we have assumed that the 
inhomogeneous distribution of the matter density can be  decomposed
into the homogeneous part, i.e., the ``background'' density and the
(small) inhomogeneous fluctuation part, i.e., the density contrast.
What is the ``background'' density in the inhomogeneous
universe?
In the actually inhomogeneous universe,
we have to operationally define the ``background'' density through the
averaging procedure.

Let us consider a finite small domain $D$ in the $t$=const. 
hypersurface $\Sigma_t$. 
The spatial volume $V$ of the domain $D$ is
\begin{equation}
  V \equiv \int_D \sqrt{\det(g_{ij})} \,d^3x.
\end{equation}
The spatial average of a spatial scalar quantity $Q$ over the
domain $D$ is in general defined by
\begin{equation}
  \langle Q \rangle \equiv \frac{1}{V} \int_D Q
  \sqrt{\det(g_{ij})} \,d^3x. 
\end{equation}
The average density in this domain is then
\begin{equation}
  \langle\rho\rangle \equiv \frac{1}{V} \int_D \rho
  \sqrt{\det(g_{ij})} \,d^3x. 
\end{equation}
The ``background'' density $\rho_b$ is obtained by averaging $\rho$
over any sufficiently large region.
Mathematically, it is \cite{3,4}
\begin{equation}
  \rho_b \equiv \lim_{D\rightarrow \Sigma_t} \langle\rho\rangle,
  \quad
  D \subset \Sigma_t . 
\end{equation}
It is assumed that this limit exists. 
Since we can observe only a finite portion of the entire universe, it is
likely that the average density $\langle\rho\rangle$ of the observed
domain $D$ is not necessarily equal to the ``background'' density
$\rho_b$,
\begin{equation}
  \langle\rho\rangle \neq \rho_b \quad\mbox{in general for $D \ll \Sigma_t$}. 
\end{equation}

From the observational point of view in the domain $D$ which is
sufficiently small compared to the entire universe, the
relevant quantity related to the cosmic expansion is not the scale
factor $a$ in the ``background'', but the domain scale
factor $\aD$ defined by the volume expansion of the domain $D$,
\begin{equation}
  3\frac{\dot{a}_{\scriptscriptstyle D}}{\aD} \equiv \frac{\dot{V}}{V}
  = \frac{1}{V}\int_D \frac{\partial}{\partial t} \sqrt{\det(g_{ij})}\, d^3x
  = \frac{1}{V}\int_D K^i_{\ i} \sqrt{\det(g_{ij})}\, d^3x =
  \langle K^i_{\ i} \rangle. 
\end{equation}

So far the treatment is exact and general.  If we use the solutions of
the linear perturbation theory
Eqs. (\ref{eq:linisol0})-(\ref{eq:linisol2}), we obtain
\begin{equation}\label{eq:aD}
  \frac{\dot{a}_{\scriptscriptstyle D}}{\aD} =
  \frac{1}{3}\left\langle K^i_{\ i}\right\rangle =
  \frac{\dot{a}}{a} + \frac{1}{3} \langle\Delta \dot{E}\rangle =
  \frac{\dot{a}}{a}\left(1 - \frac{1}{3}
    \langle\delta\rangle\right). 
\end{equation}

The Friedmann equation for $\aD$ can be obtained by spatially
averaging the Einstein equation Eq.~(\ref{eq:G00}).
Again, if we use the linear order solutions, we obtain, up to the linear
order of the perturbations, 
\begin{equation}\label{eq:aD2}
  \left(\frac{\dot{a}_{\scriptscriptstyle D}}{\aD}\right)^2 +
  \frac{K_{\scriptscriptstyle \mbox{\tiny
        eff}}}{a^2_{\scriptscriptstyle D}}
  = \frac{8\pi G}{3} \langle\rho\rangle, 
\end{equation}
where
\begin{equation}
  K_{\mbox{\tiny eff}} \equiv
  -\frac{2}{3}\langle \Delta F \rangle = 
  \frac{10}{9}
  \left\langle\Delta\phi(\v{x})\right\rangle \propto
  \left\langle \delta \right\rangle
\end{equation}
is a constant which can be regarded as the effective curvature
constant in the domain $D$. 

It should be emphasized that observed part of the domain $D$, which is
weakly inhomogeneous, may behave on average as if it were of constant
curvature with $K_{\mbox{\tiny eff}}\neq 0$, even if we have assumed that
the ``background'' is spatially flat, $K=0$.
If $\left\langle\delta\right\rangle > 0$, then
$K_{\mbox{\tiny eff}} >0$, and
if  $\left\langle\delta\right\rangle < 0$, then
$K_{\mbox{\tiny eff}} <0$.  

The energy equation for the domain average density $\langle\rho\rangle$ is
obtained by averaging Eq.~(\ref{eq:ene}):
\begin{equation}\label{eq:eneD}
  \frac{d}{dt}\langle\rho\rangle + 3\frac{\dot{a}_{\scriptscriptstyle D}}{\aD}
  \langle\rho\rangle = 0. 
\end{equation}
Note that Eq.~(\ref{eq:eneD}) holds exactly without any approximation,
thanks to the following commutation rule
\begin{equation}
  \left\langle\frac{\partial}{\partial t} Q \right\rangle
  -\frac{d}{dt}\left\langle Q\right\rangle =
  \left\langle K^i_{\ i}\right\rangle\left\langle
    Q\right\rangle
  -\left\langle K^i_{\ i} \,Q\right\rangle .
\end{equation}

\section{The cosmological parameters
  measured in the nearby regions}

In spite of the recent progress in observational technology,
we can still observe only a finite part of the domain $D$,
which is still small compared to the entire universe.
Therefore,  the domain average density $\langle\rho\rangle$ plays the
important role to drive  the cosmic expansion of the
observed domain  of volume $V$.
The cosmological parameters which are determined from  the observations in
the nearby regions may not be necessarily equal to those in the
``background'' universe.
Let us clarify this situation.

We define the global Hubble parameter $H_0$ by
\begin{equation}
  H_0\equiv \frac{\dot{a}}{a}\biggr|_{t_0}, 
\end{equation}
and the global density parameter $\Omega_0$ by
\begin{equation}
  \Omega_0 \equiv \frac{8\pi G\rho_b(t_0)}{3 H^2_0}, 
\end{equation}
which is unity since we have assumed the $K=0$ background. 

On the other hand, the cosmological parameters determined from the
local observations in nearby regions of volume $V$, which are sufficiently
small compared to the entire universe,
are certainly characterized by $\aD$ which is driven by the average
density 
$\langle\rho\rangle$ in this region.

Therefore, it is natural to define the local Hubble parameter
$\tilde{H}_0$ by
\begin{equation}
  \tilde{H}_0 \equiv \frac{\dot{a}_{\scriptscriptstyle D}}{\aD}\biggr|_{t_0}, 
\end{equation}
and the local density parameter $\tilde{\Omega}_0$ by
\begin{equation}\label{eq:tilom}
  \tilde{\Omega}_0 \equiv \frac{8\pi G \langle\rho(t_0)\rangle}{3
    \tilde{H}_0^2}. 
\end{equation}
From Eqs.~(\ref{eq:aD}) and (\ref{eq:tilom}), we obtain the relation
between the local and the global cosmological parameters as
\begin{equation}
  \tilde{H}_0 = H_0 \left(1 - \frac{1}{3}\langle\delta\rangle_{t_0} \right),
\end{equation}
\begin{equation}
  \tilde{\Omega}_0 =
  \frac{8\pi G\rho_b(t_0)\left(1+\langle\delta\rangle_{t_0}\right)}
       {3 H^2_0\left(1-\frac{1}{3}\langle\delta\rangle_{t_0}\right)^2}=
  \Omega_0\left(
    1+\frac{5}{3}\langle\delta\rangle_{t_0} \right),  
\end{equation}
up to the linear order of the density perturbation $\delta$.

The local cosmological parameters coincide with the global ones if and
only if $\langle\delta\rangle =0$, i.e., $\langle\rho\rangle =
\rho_b$.
A rough estimation shows that a 30\% under-dense region, i.e.,
$\langle\delta\rangle_{t_0} = -0.3$ can explain the 10\% larger value of
the local Hubble parameter $\tilde{H}_0$ compared to the
global $H_0$.

It should also be noted that the density parameter may change the
value in different measurements in different scales.
For example, 
if the local Hubble parameter has a higher value than that of
the global one, $\tilde{H}_0 > H_0$,
then, the local region has a lower density parameter,
$\tilde{\Omega}_0 < \Omega_0$. 

\section{Conclusion}

We have operationally defined the average behavior of the actual,
inhomogeneous universe.
Since the observed region is finite and sufficiently
small compared to the entire universe, the cosmic expansion of this region is
driven by the domain average density $\langle\rho\rangle$, the spatial averaging of the
inhomogeneous distribution of matter over this finite region, which is
not always coincident with the ``background'' density $\rho_b$.

We have also shown that the cosmological parameters determined by the
local observations in a finite nearby regions may differ from the
large-scale, ``background'' ones, which may be helpful towards
solving  the Hubble constant problem.
In particular, about 10\% difference between the local and the
global Hubble parameters may by safely explained within the
framework of linear perturbation theory, with the help of spatial averaging
procedure defined over a finite spatial domain in the
$t=\mbox{const.}$ hypersurface. 

Finally, we would like to mention an interesting possibility of
solving apparent acceleration of the cosmic expansion. 
One of the present authors has re-analyzed the observed magnitude-redshift
($m$-$z$)
relation of type Ia supernovae (SNe Ia) and has examined the
possibility that the  
apparent acceleration of the cosmic expansion is not caused by dark
energy by is instead of a consequence of the large-scale
inhomogeneities in the universe \cite{5}.
He has concluded that, assuming the inhomogeneous Hubble parameter,   
a larger value of $H_0$  in the nearby, low-redshift region than that in the 
distant, high-redshift region  may be
sufficient to explain the observed $m$-$z$ relation for SNe Ia, without
introducing dark energy.
At that time, the author has proposed only a phenomenological
description of the large-scale inhomogeneities, and has not given
a physical explanation why the Hubble parameter can change
between the nearby and distant regions.

Now we have a plausible explanation: the value of the local Hubble
parameter $\tilde{H}_0$ 
may be different from that of the global one $H_0$, if the
domain average density $\langle\rho\rangle$ in the locally observed  region
is different from the  ``background'' one $\rho_b$.

Therefore, we hope that the idea proposed in this paper 
may give a simple and interesting tool towards resolving not only the
Hubble parameter discrepancy but also the apparent acceleration of the
cosmic expansion mystery.

\section*{Acknowledgment}

We would like to thank K.~Hasegawa and K.~Morimoto
for the helpful discussions in Hirosaki University.

\end{document}